\documentstyle[sprocl,psfig]{article}

\newcommand{\msun}{M_{\odot}}

\newcommand{\eos}{equation of state~}

\newcommand{\eosp}{equation of state}

\newcommand{\smh}{strange matter hypothesis~}

\newcommand{\beqn}{\begin{eqnarray}}
\newcommand{\eeqn}{\end{eqnarray}}
\newcommand{\okgr}{\Omega_{\rm K}}

\newcommand{\pkgr}{P_{\rm K}}

\newcommand{\msec}{\rm msec}

\newcommand{\gsim}{\stackrel{\textstyle >}{_\sim}}
\newcommand{\lsim}{\stackrel{\textstyle <}{_\sim}}

\newcommand{\gcmt}{{\rm g/cm}^3}

\newcommand{\edrip}{\epsilon_{\rm drip}}

\newcommand{\ecrusti}{\epsilon_{\rm crust}}

\def\be{\begin{equation}}
\def\ee{\end{equation}}
\def\bea{\begin{eqnarray}}
\def\eea{\end{eqnarray}}

\begin{document}

\fbox{\parbox[t]{7cm}{\scriptsize To be published in the proceedings of the
International Conference on Nuclear Physics at the Turn of the
Millennium: Structure of Vacuum and Elementary Matter, March 10-16,
1996, Wilderness, South Africa (World Scientific 1996)}}
\bigskip

\title{QUARK MATTER, MASSIVE STARS AND STRANGE PLANETS}

\author{F. WEBER, CH. SCHAAB, M. K. WEIGEL}

\address{Institute for Theoretical Physics, University of Munich,\\
Theresienstr. 37, 80333 Munich, Germany}

\author{N. K. GLENDENNING}

\address{Nuclear Science Division \& Institute for Nuclear and Particle \\
Astrophysics, Lawrence Berkeley National Laboratory, \\
Berkeley, CA 94720, USA}

\maketitle\abstracts{This paper gives an overview of the properties of all 
  possible equilibrium sequences of compact strange-matter stars with
  nuclear crusts, which range from strange stars to strange dwarfs. In
  contrast to their non-strange counterparts, --neutron stars and
  white dwarfs--, their properties are determined by two (rather than
  one) parameters, the central star density and the density at the
  base of the nuclear crust. This leads to stellar strange-matter
  configurations whose properties are much more complex than those of
  the conventional sequence.  As an example, two generically different
  categories of stable strange dwarfs are found, which could be the
  observed white dwarfs.  Furthermore we find very low-mass strange
  stellar objects, with masses as small as those of Jupiter or even
  lighter planets.  Such objects, if abundant enough in our Galaxy,
  should be seen by the presently performed gravitational microlensing
  searches. Further aspects studied in this paper concern the limiting
  rotational periods and the cooling behavior of neutron stars and
  their strange counterparts.}

\section{Introduction}

The theoretical possibility that strange quark matter may be
absolutely stable with respect to iron (energy per baryon below 930
MeV) has been pointed out by Bodmer\,\cite{bodmer71:a},
Witten\,\cite{witten84:a}, and Terazawa\,\cite{terazawa89:a}.  This
so-called strange matter hypothesis constitutes one of the most
startling possibilities of the behavior of superdense nuclear matter,
which, if true, would have implications of greatest importance for
cosmology, the early universe, its
\begin{figure}[tb]
\begin{center}
\leavevmode
\psfig{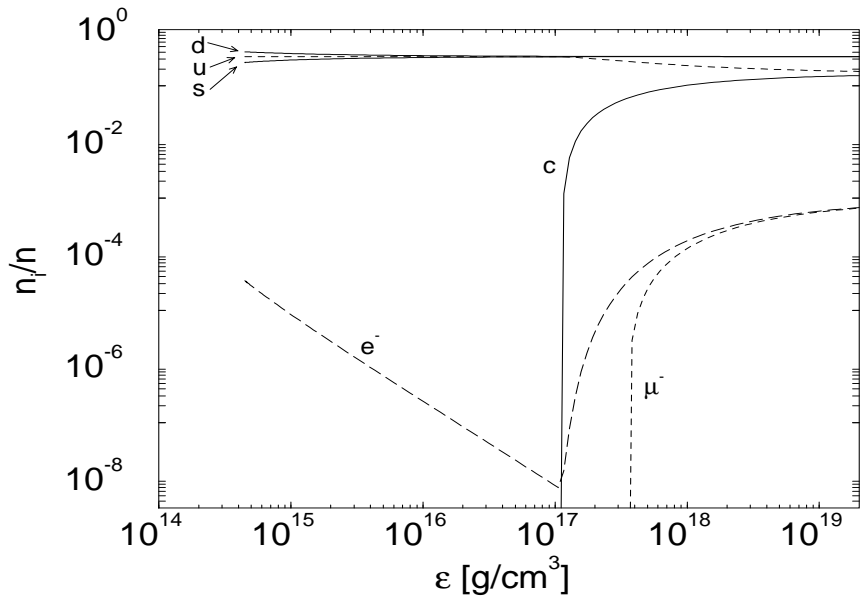}
\caption[Relative densities of quarks and leptons in cold, beta-stable, 
electrically charge neutral quark matter versus density.]{Relative
densities of quarks and leptons in absolutely stable
strange-quark-star matter versus density. $n_i$ and $n$ denote partial
and total densities, respectively.}
\label{fig:1.5}
\end{center}
\end{figure}
evolution to the present day, astrophysical compact objects, and
laboratory physics\,\cite{aarhus91:proc}.  Unfortunately it seems
unlikely that QCD calculations will be accurate enough in the
foreseeable future to give a definitive prediction on the absolute
stability of strange matter, such that one is left with experiments
and astrophysical tests, as performed here, to either confirm or
reject the hypothesis.

One striking implication of the hypothesis would be that pulsars,
which are conventionally interpreted as rotating neutron stars, almost
certainly would be rotating strange stars (strange pulsars). Part of
this paper deals with an investigation of the properties of such
objects.  In
\begin{figure}[tb]
\begin{center}
\leavevmode
\psfig{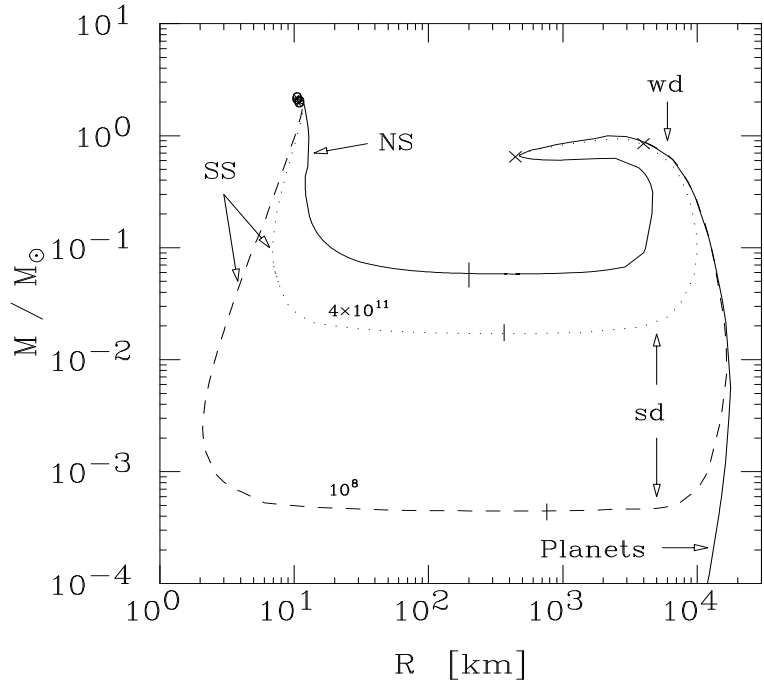}
\caption[Neutron star--white dwarf sequences and their strange 
counterparts]{Neutron star (NS)--white dwarf (wd) sequence (solid
line). The dotted and dashed curves refer to strange star
(SS)--strange dwarf (sd) sequences with inner crust densities as
indicated (in $\gcmt$).  Vertical bars mark minimum mass stars,
crosses mark the termination points of the strange star
sequences.}
\label{fig:sequence}
\end{center}
\end{figure}
addition to this, we develop the complete sequence of strange stars
with nuclear crusts, which ranges from the compact members, with
properties similar to those of neutron stars, to white-dwarf-like
objects (strange dwarfs) and discuss their stability against
acoustical vibrations\,\cite{weber93:b}.  The
properties with respect to which strange-matter stars differ from
their non-strange counterparts are discussed.

\goodbreak
\section{Quark-lepton Composition of Strange Matter}\label{sec:qlc}

The relative quark--lepton composition of quark-star matter at zero
temperature is shown in Fig.\
\ref{fig:1.5}\,\protect{\cite{kettner94:b}.  All quark flavor
states that become populated at the densities shown are taken into
account.  Since stars in their lowest energy state are electrically
charge neutral to very high precision, any net positive quark charge
must be balanced by leptons. In general, as can be seen in Fig.\
\ref{fig:1.5}, there is only little need for leptons, since charge
neutrality can be achieved essentially among the quarks themselves.
The concentration of electrons is largest at the lower densities of
Fig.\ \ref{fig:1.5}, due to the finite strange-quark mass which leads
to a deficit of net negative quark charge, and at densities beyond
which the charm-quark state becomes populated which increases the net
positive quark charge.

\goodbreak
\section{Nuclear Crusts on Strange Stars and Equation of State}

The presence of electrons in strange quark matter is crucial for the
possible existence of a nuclear crust on such objects.  As shown in
Refs.\ \cite{kettner94:b,alcock86:a}, the electrons, because they are
bound to strange matter by the Coulomb force rather than the strong
force, extend several hundred fermi beyond the surface of the strange
star.  Associated with this electron displacement is a very strong
electric dipole layer which can support -- out of contact with the
surface of the strange star -- a crust of nuclear material, which it
polarizes.  The maximal possible density at the base of the crust
(inner crust density) is determined by the neutron drip density
($\edrip=4.3\times 10^{11}~\gcmt$), at which neutrons begin to drip out
of the nuclei and form a free neutron gas. Being electrically charge
neutral, the neutrons do not feel the repulsive Coulomb force and
hence would gravitate toward the quark matter core, where they become
converted, via hypothesis, into strange quark matter. So neutron drip
sets a strict upper limit on the crust's maximal inner density.

The somewhat complicated situation of the structure of a strange star
with crust can be represented by a proper choice of
\eosp\,\cite{glen92:crust}, which consists of two parts. At densities
below neutron drip it is represented by the low-density \eos of
charge-neutral nuclear matter, for which we use the
Baym-Pethick-Sutherland \eosp.  The star's strange-matter core is
described by the bag model. The graphical illustration of such an \eos
can be found in Ref.\ \cite{glen92:crust}.  

\goodbreak
\section{Complete Sequences of Strange-Matter Stars}

Since the nuclear crusts surrounding the cores of strange stars are
bound by the gravitational force rather than confinement, the
mass-radius relationship of strange-matter stars with crusts is
qualitatively similar to the one of purely gravitationally bound
stars --neutron stars and white dwarfs--, as illustrated in Fig.\
\ref{fig:sequence}\,\protect{\cite{weber93:b,glen94:a}.  
The strange-star sequences are computed for the maximal possible inner
crust density, $\ecrusti=\edrip$, as well as for an arbitrarily
chosen, smaller value of $\ecrusti=10^8~\gcmt$, which may serves to
demonstrate the influence of less dense crusts on the mass-radius
relationship\,\cite{weber93:b}.  From the maximum-mass star (dot), the
central density decreases monotonically through the sequence in each
case.  The neutron-star sequence is computed for a representative
model for the \eos of neutron star matter, the relativistic
Hartree-Fock \eos of Ref.\ \cite{weber89:e}, which has been combined
at subnuclear densities with the Baym-Pethick-Sutherland
\eosp.  Hence the white dwarfs shown in Fig.\ \ref{fig:sequence} are
computed for the latter.  The gravitationally bound stars with radii
$\lsim 200$ km or $\gsim 3000$ km represent stable neutron stars and
white dwarfs, respectively. 

The fact that strange stars with crusts tend to possess somewhat
smaller radii than neutron stars leads to smaller rotational mass
shedding (Kepler) periods $\pkgr$ for the former, as is indicated
classically by $\pkgr=2\pi\sqrt{R^3/M}$. Of course the general
relativistic expression, 
\begin{equation} 
\pkgr \equiv 2 \pi/\okgr \, , ~ {\rm with}~~~
\okgr = \omega + \frac{\omega^\prime}{2\psi^\prime} +e^{\nu -\psi} \sqrt{
\frac{\nu^\prime}{\psi^\prime} + \Bigl(\frac{\omega^\prime}{2
\psi^\prime}e^{\psi-\nu}\Bigr)^2 } \, ,  \label{eq:okgr}
\end{equation} 
which is to be applied to neutron and strange stars, is considerably
more complicated. It is to be computed simultaneously in combination
with Einstein's equations (see Ref.\ \cite{weber93:b} for details and
further references),
\begin{equation} 
 {\cal R}^{\kappa\lambda} \; - \; {1\over 2} \; g^{\kappa\lambda} \;
 {\cal R} \; = \; 8\, \pi \; {\cal
 T}^{\kappa\lambda}(\epsilon,P(\epsilon)) \; , \label{eq:einstein}
\end{equation}
However the qualitative dependence of $\pkgr$ on mass and radius as
expressed by the classical expression remains valid. So one finds
that, due to the smaller radii of strange stars, the complete
sequence of such objects (and not just those close to the mass peak,
as is the case for neutron stars) can sustain extremely rapid
rotation\,\cite{weber93:b}.  In particular, our model calculations
indicate for a strange star with a typical pulsar mass of $\sim
1.45\,\msun$  Kepler periods as small as $0.55\lsim\pkgr/\msec\lsim
0.8$, depending on crust thickness and bag
constant\,\cite{glen92:crust,weber93:b}. This range is to be compared
with $\pkgr\sim 1~\msec$ obtained for neutron stars of the same
mass.

The minimum-mass configurations of the strange-star sequences
in Fig.\ \ref{fig:sequence} have masses of about $\sim 0.017\,
\msun$ (about 17 Jupiter masses) and  $10^{-4}\,\msun$, depending 
on inner crust density.  More than that, for inner crust densities
smaller than $10^8~\gcmt$ we find strange-matter stars that can be even
by orders of magnitude lighter than Jupiters.  If abundant enough in
our Galaxy, all these light strange stars could be seen by the
gravitational microlensing searches that are being performed
presently.  Strange stars located to the right of the minimum-mass
configuration of each sequence consist of small strange cores ($\lsim
3$ km) surrounded by a thick nuclear crust, made up of white dwarf
material.  We thus call such objects strange dwarfs.  Their cores have
shrunk to zero at the crossed points. What is left is an ordinary
white dwarf with a central density equal to the inner crust density of
the former strange dwarf\,\cite{weber93:b,glen94:a}.  A detailed
stability analysis of strange stars against radial
oscillations\,\cite{weber93:b} shows that all those strange-dwarf
sequences that terminate at stable ordinary white dwarfs are stable
against radial oscillations.  Strange stars that are located to the
left of the mass peak of ordinary white dwarfs, however, are unstable
against oscillations and thus cannot exist stably in nature.  So, in
sharp contrast to neutron stars and white dwarfs, the branches of
strange stars and strange dwarfs are stably connected with each
other\,\cite{weber93:b,glen94:a}.

Finally the strange dwarfs with $10^9~\gcmt<\ecrusti<4\times
10^{11}~\gcmt$ form entire new classes of stars that contain nuclear
material up to $\sim 4\times 10^4$ times denser than in ordinary white
dwarfs of average mass, $M\sim 0.6\,\msun$ (central density $\sim
10^7~\gcmt$).  The entire family of such strange stars owes its
stability to the strange core.  Without the core they would be placed
into the unstable region between ordinary white dwarfs and neutron
stars\,\cite{glen94:a}.

\section{Thermal Evolution of Neutron Stars and Strange Stars}

The left panel of Fig. \ref{fig:cool} shows a numerical simulation of
the thermal evolution of neutron stars.  The neutrino emission rates
are determined by the modified and direct Urca processes, and the
presence of a pion or kaon condensate.  The baryons are treated as
superfluid particles. Hence the neutrino emissivities are suppressed
by an exponential factor of $\exp(-\Delta/kT)$, where $\Delta$ is the
width of the superfluid gap (see Ref.\, \cite{schaab95:a} for
details).  Due to the dependence of the direct Urca process and the
onset of meson condensation on star mass, stars that are too light for
these processes to occur (i.e., $M<1\,\msun$) are restricted to
standard cooling via modified Urca.  Enhanced cooling via the other
three processes results
\begin{figure}[tb] \centering 
\leavevmode
\psfig{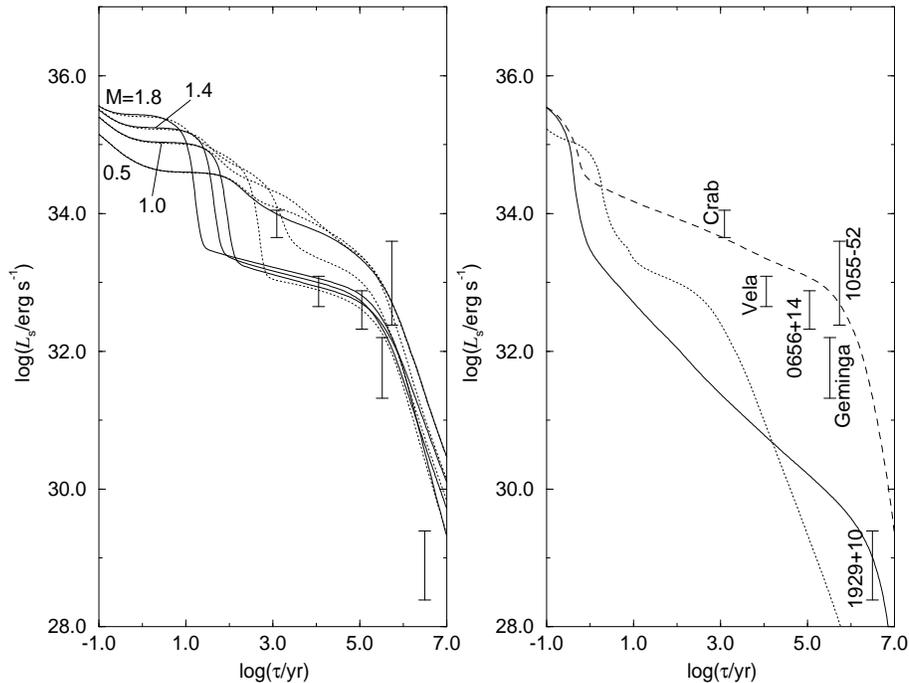} 
  \caption{Left panel: Cooling of neutron stars with pion (solid
  curves) or kaon condensates (dotted curve). Right panel: Cooling of
  $M=1.8\,M_\odot$ strange stars with crust. The cooling curves of
  lighter strange stars, e.g. $M\gsim 1\,M_\odot$, differ only
  insignificantly from those shown here. Three different assumptions
  about a possible superfluid behavior of strange quark matter are
  made: no superfluidity (solid), superfluidity of all three flavors
  (dotted), and superfluidity of up and down flavors only
  (dashed). The vertical bars denote luminosities of observed pulsars.
  \label{fig:cool}}
\end{figure}
in a sudden drop of the star's surface temperature after about 10 to
$10^3$ years after birth, depending on the thickness of the ionic
crust.  As one sees, agreement with the observed data is achieved
only if different masses for the underlying pulsars are assumed.
The right panel of Fig. \ref{fig:cool} shows cooling simulations of
strange quark stars. The curves differ with respect to assumptions
made about a possible superfluid behavior of the quarks. Because of
the higher neutrino emission rate in non-superfluid quark matter, such
quark stars cool most rapidly (as long as cooling is core
dominated). In this case one does not get agreement with most of the
observed pulsar data. The only exception is pulsar PSR
1929+10. Superfluidity among the quarks reduces the neutrino emission
rate, which delays cooling\,\cite{schaab95:a}. This moves the cooling
curves into the region where most of the observed data lie.

Subject to the inherent uncertainties in the behavior of strange quark
matter as well as superdense nuclear matter, at present it appears
much too premature to draw any definitive conclusions about the true
nature of observed pulsars. Nevertheless, should a continued future
analysis in fact confirm a considerably faster cooling of strange
stars relative to neutron stars, this would provide a definitive
signature (together with rapid rotation) for the identification of a
strange star. Specifically, the prompt drop in temperature at the very
early stages of a pulsar, say within the first 10 to 50 years after
its formation, could offer a good signature of strange
stars\,\cite{pizzochero91:a}.  This feature, provided it withstands a
more rigorous analysis of the microscopic properties of quark matter,
could become particularly interesting if continued observation of SN
1987A would reveal the temperature of the possibly existing pulsar at
its center.

\section{Summary}

This work deals with an investigation of the properties of the
complete sequences of strange-matter stars that carry nuclear crusts.
The following items are particularly noteworthy:

\begin{enumerate}
\item The complete sequence of compact strange stars can sustain
  extremely rapid rotation and not just those close to the mass peak,
  as is the case for neutron stars!

\item If the \smh is correct, the observed white dwarfs 
  and planets could contain strange-matter cores in their centers. The
  baryon numbers of their cores are smaller than $\lsim 2 \times
  10^{55}$!

\item The strange stellar configurations would populate a vast region in 
  the mass-radius plane of collapsed stars that is entirely void
  of stars if strange quark matter is not the absolute ground state!

\item If the new classes of stars mentioned in (2) and (3) exist
  abundantly enough in our Galaxy, the presently performed
  gravitational microlensing experiments could see them all!

\item Due to the uncertainties in the behavior of superdense
  nuclear as well as strange matter, no definitive conclusions about
  the true nature (strange or conventional) of observed pulsar can be
  drawn from cooling simulations yet.  As of yet they could be made of
  strange quark matter as well as of conventional nuclear matter.
\end{enumerate}

Of course, there remain various interesting aspects of strange
pulsars, strange dwarfs and strange planets, that need to be worked
out in detail. From their analysis one may hope to arrive at
definitive conclusion about the behavior of superdense nuclear matter
and, specifically, the true ground state of strongly interacting
matter.

\end{document}